\documentclass[11pt]{article}
\usepackage{epsf,amsmath,amssymb}
\newcommand{\abbrev}{\small}
\newcommand{\ep}{\epsilon}
\newcommand{\api}{\frac{\alpha_s}{\pi}}
\newcommand{\eqn}[1]{Eq.\,(\ref{#1})}
\newcommand{\fig}[1]{Fig.\,\ref{#1}}

\newcommand{\dd}{{\rm d}}
\newcommand{\ddoverdd}[1]{\frac{\dd}{\dd #1}}

\newcommand{\order}[1]{{\cal O}(#1)}
\newcommand{\lo}{{\abbrev LO}}
\newcommand{\nlo}{{\abbrev NLO}}
\newcommand{\nnlo}{{\abbrev NNLO}}

\newcommand{\reference}[1]{Ref.\,\cite{#1}}

\newcommand{\dred}{{\abbrev DRED}}
\newcommand{\dreg}{{\abbrev DREG}}

\newcommand{\Msusy}[1]{\tilde M_{#1}}

\newcommand{\Mtop}{M_t}
\newcommand{\Mstop}{\Msusy{t}}
\newcommand{\Mgluino}{\Msusy{g}}
\newcommand{\Ltop}{L_t}

\newcommand{\LstopL}{L_L}
\newcommand{\LstopR}{L_R}
\newcommand{\Lgluino}{L_g}
\newcommand{\sm}{{\abbrev SM}}
\newcommand{\mssm}{{\abbrev MSSM}}
\newcommand{\susy}{{\abbrev SUSY}}
\newcommand{\higgs}{\phi}
\newcommand{\lhc}{{\abbrev LHC}}

\newcommand{\Mts}{x_{ts}}
\newcommand{\MtL}{x_{tL}}
\newcommand{\MLg}{x_{Lg}}
\newcommand{\MtR}{x_{tR}}
\newcommand{\Mtg}{x_{tg}}

\newcommand{\LgR}{L_{gR}}
\newcommand{\LgL}{L_{gL}}
\newcommand{\Ltg}{L_{tg}}
\newcommand{\Lgs}{L_{gs}}
\renewcommand{\thefootnote}{\fnsymbol{footnote}}
\begin{document}    

\title{\vskip-3cm{\baselineskip14pt
\centerline{\normalsize\hfill CERN-TH 2003-155}
\centerline{\normalsize\hfill DESY 03--084}
\centerline{\normalsize\hfill hep-ph/0307346}
\centerline{\normalsize\hfill July 2003}
}
\vskip.7cm
Hadronic Higgs Production and Decay in Supersymmetry at
Next-to-Leading Order
\vskip.3cm
}
\author{
{Robert V. Harlander}$^{(a)}$
\,and
{Matthias Steinhauser}$^{(b)}$
  \\[3em]
  {\it (a) Theory Division, CERN, CH-1211 Geneva 23, Switzerland}\\
    {{\it email:} \tt robert.harlander@cern.ch}\\
  \\[.5em]
  {\it (b) II. Institut f\"ur Theoretische Physik,}\\ 
  {\it Universit\"at Hamburg, D-22761 Hamburg, Germany}\\
  {{\it email:} \tt matthias.steinhauser@desy.de}
}
\date{}
\maketitle

\begin{abstract}
\noindent
Supersymmetric {\abbrev QCD} corrections to the gluonic production and
decay rate of a CP-even Higgs boson are evaluated at next-to-leading
order.  To this aim, we derive an effective Lagrangian for the
gluon-Higgs coupling.  We show that a consistent calculation requires
the inclusion of gluino effects, in contrast to what has been done
previously.  The supersymmetric corrections to the gluon-Higgs coupling
lead to a modification of the next-to-leading order K-factor for the
Higgs production rate at the {\abbrev LHC} by less than 5\%.
\end{abstract}

\renewcommand{\thefootnote}{\arabic{footnote}}
\setcounter{footnote}{0}

\section{Introduction}

It is among the main goals of the {\abbrev CERN} Large Hadron Collider
({\abbrev LHC}) to observe the Higgs boson and study its properties.
The most important production mechanisms for a Standard Model (\sm{})
Higgs boson have been studied in great detail and are known with fairly
high theoretical precision (for a review, 
see Refs.~\cite{Kniehl:1993ay,Spira:1997dg,Carena:2002es}). The
dominant mode is gluon fusion, where, at leading order in perturbation
theory, the gluons couple to the Higgs boson preferably via a top quark loop.
A light Higgs boson, $M_H\leq 130$\,GeV, would be observed through its
subsequent decay into photon pairs.  For larger values of the Higgs
boson mass, $H\to ZZ^{(\ast)}$ and $H\to WW^{(\ast)}$ allow for a clear
identification of the Higgs signal.

The next-to-leading order (\nlo{}) corrections to the gluon fusion
process in the Standard Model are known
exactly~\cite{Graudenz:1992pv,Spira:1995rr,Spira:1997dg}. They were
shown to be approximated extremely well by an earlier calculation which
was based on the heavy-top limit~\cite{Dawson:1990zj,Djouadi:1991tk}.
This is true for a large range of the Higgs mass, including even
$M_H>2\,\Mtop{}$, provided that the mass dependence at leading order in
$\alpha_s$ is kept exactly.  Recently, the next-to-next-to-leading order
(\nnlo{}) corrections in the heavy-top limit became
available~\cite{Harlander:2002wh,Anastasiou:2002yz} (see also
\reference{Ravindran:2003um}). They lead to a significant stabilization
of the theoretical prediction, with a remaining scale uncertainty of
around 15\%. The \sm{} Higgs production rate at the \lhc{} is thus under
good theoretical control.

In the Minimal Supersymmetric Standard Model (\mssm), several aspects
have to be taken into account when considering the production of a
neutral scalar Higgs boson: For example, for large $\tan\beta$, the
bottom quark can contribute significantly to the gluon-Higgs coupling.
This effect has been studied at \nlo{} in
Refs.~\cite{Graudenz:1992pv,Spira:1993bb,Spira:1995rr,Spira:1997dg}.
A large value of
$\tan\beta$ also brings in the associated production of a Higgs boson
with bottom quarks as one of the main production
mechanisms~\cite{Eichten:1984eu,Maltoni:2003pn,Harlander:2003ai}.  

The main subject of the current paper are the contributions of \mssm{}
particles to the gluon-Higgs coupling. At sufficiently small values of
$\tan\beta$, the dominant effects are due to top-squarks and gluinos.
The \nlo{} squark contributions have been addressed in
\reference{Dawson:1996xz}.  However, as we will show, these results do
not correspond to a consistent supersymmetric framework. In fact, it
turns out that the squark contributions can not be separated from the
gluino effects.  This will significantly affect the final result, most
evidently in the appearance of non-decoupling logarithmic contributions
from the gluino mass.

Motivated by its success in the \sm{} case, we construct an
effective Lagrangian at \nlo{} for the gluon-Higgs interaction in the
\mssm{}. This corresponds to integrating out the top quark and all
supersymmetric particles. Within the framework of this effective
Lagrangian, all calculations can be performed in standard five-flavor
{\abbrev QCD}. They can thus be taken over from the \sm{} case.

In this paper, we restrict ourselves to the case without squark mixing,
and focus on a principle study of the method.  A more general analysis
will be presented in a forthcoming paper.

\section{The effective Lagrangian in supersymmetric QCD}\label{sec::efflag}
\subsection{General definition}
It has been shown that the radiative corrections to the production rate
of a \sm{} Higgs boson are well described by an effective theory, where
{\it formally} the top quark is assumed much heavier than the Higgs
boson. In practice, it turns out that this approximation works very well
even way beyond the formal limit $\Mtop{}>M_H$.  We therefore assume that a
similar behavior will be observed in the \mssm{}, and construct an
effective Lagrangian where the top quark, the squarks, and the gluino
are integrated out.  The Higgs boson itself enters only as external
particle and thus the Lagrangian describing the effective $\higgs gg$ coupling
at leading order in $M_\higgs$ has the form
\begin{eqnarray}
  {\cal L}_{Y,\rm eff} &=& -\frac{\higgs^0}{v^0} C_1^0 {\cal O}_1^0
  \,,\qquad {\cal O}_1^0 = \frac{1}{4} G_{\mu\nu}^0 G^{0,\mu\nu}\,,
  \label{eq::leff}
\end{eqnarray}
where $C_1^0$ is the coefficient function containing the remnant
contribution of the heavy particles and ${\cal O}_1^0$ is the effective
operator. $\phi$ denotes either of the two {\abbrev CP}-even Higgs
bosons of the \mssm{}.  $G_{\mu\nu}^0$ is the gluonic field strength
tensor of standard, five-flavor {\abbrev QCD}.  The superscript ``0''
indicates bare quantities as opposed to renormalized ones.  Note that
the renormalization of $\phi^0/v^0$ is of higher order in the
electromagnetic coupling constant.

The renormalization of $C_1^0$ and ${\cal O}_1^0$ is discussed
in~\reference{Spiridonov:br,Chetyrkin:1997un} 
and is given by
\begin{equation}
\begin{split}
C_1 &= Z_{11}^{-1}\,C_1^0\,,\qquad
{\cal O}_1 = Z_{11}\,{\cal O}_1^0\,,\qquad
Z_{11} = \left(1 - \frac{\pi}{\alpha_s}\frac{\beta}{\ep}\right)^{-1}\,,
\label{eq::c1ren}
\end{split}
\end{equation}
where $\beta(\alpha_s)$ is the $\beta$-function of standard
$(n_l=5)$-flavor {\abbrev QCD}:
\begin{equation}
\begin{split}
\mu^2\ddoverdd{\mu^2} \frac{\alpha_s}{\pi} = \beta(\alpha_s)\,,
\end{split}
\end{equation}
with $\beta(\alpha_s)=-(\alpha_s/\pi)^2\beta_0+\ldots$ and
$\beta_0=11/4-n_l/6$.  Note that here and in what follows, $\alpha_s$
denotes the strong coupling constant in standard five-flavor {\abbrev QCD};
it is a function of the renormalization scale $\mu$:
\begin{equation}
\begin{split}
\alpha_s \equiv \alpha_s^{(5)}(\mu)\,.
\label{eq::defalphas}
\end{split}
\end{equation}

According to \eqn{eq::leff}, the calculation of the gluonic Higgs production
cross section or the decay rate splits into two parts: On the one hand, the
coefficient
function $C_1$ has to be determined to the desired order. This part is
independent of the process considered; the result will be useful for any
process to which the effective Lagrangian approach is applicable.  The
second part of the calculation is to evaluate the production or decay
rate on the
basis of the operator ${\cal O}_1$ within five-flavour {\abbrev QCD}.

\begin{figure}
  \begin{center}
    \leavevmode
    \begin{tabular}{cc}
      \epsfxsize=9em
      \raisebox{0em}{\epsffile[120 500 380 680]{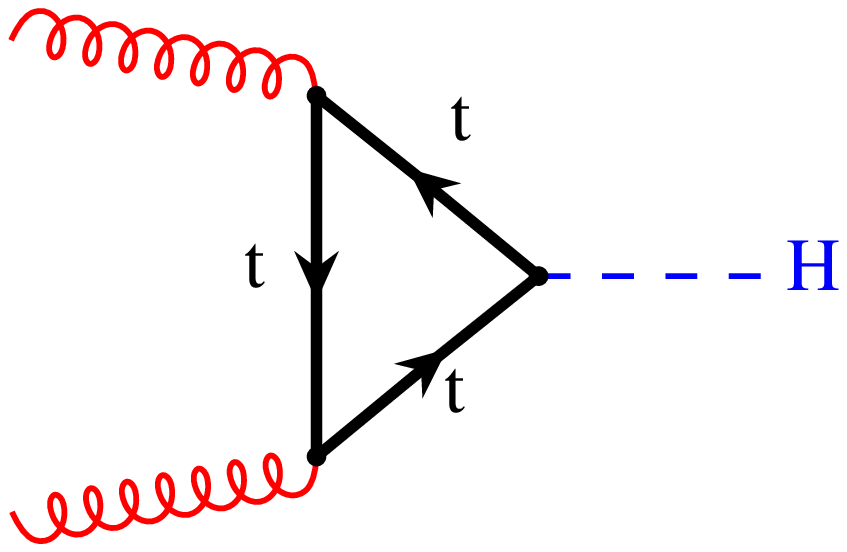}} &
      \epsfxsize=9em
      \raisebox{0em}{\epsffile[120 500 380 680]{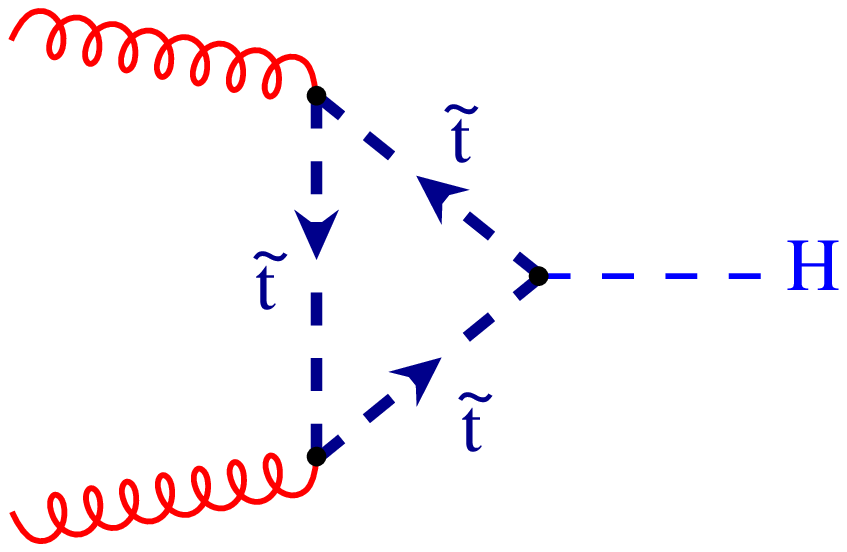}} \\
      $(a)$ & $(b)$
    \end{tabular}
    \parbox{14.cm}{
      \caption[]{\label{fig::vertlo}\sloppy
        One-loop Feynman diagrams contributing to gluon fusion in the \mssm{}.
        }}
  \end{center}
\end{figure}

\begin{figure}
  \begin{center}
    \leavevmode
    \begin{tabular}{cccc}
      \epsfxsize=9em
      \raisebox{0em}{\epsffile[120 500 380 680]{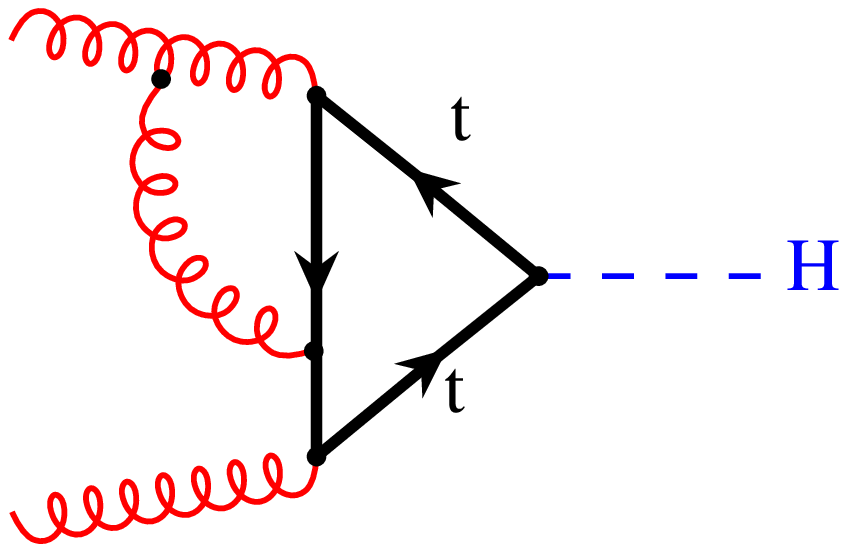}} &
      \epsfxsize=9em
      \raisebox{0em}{\epsffile[120 500 380 680]{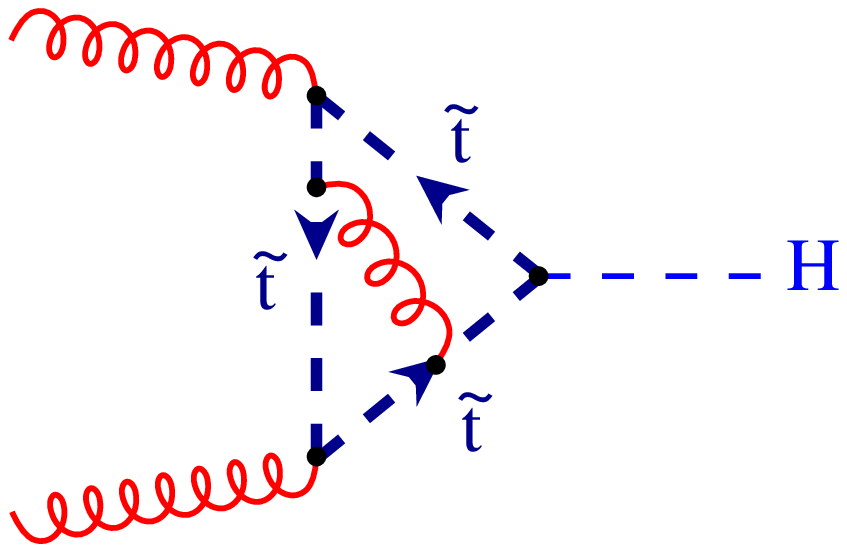}} &
      \epsfxsize=9em
      \raisebox{0em}{\epsffile[120 500 380 680]{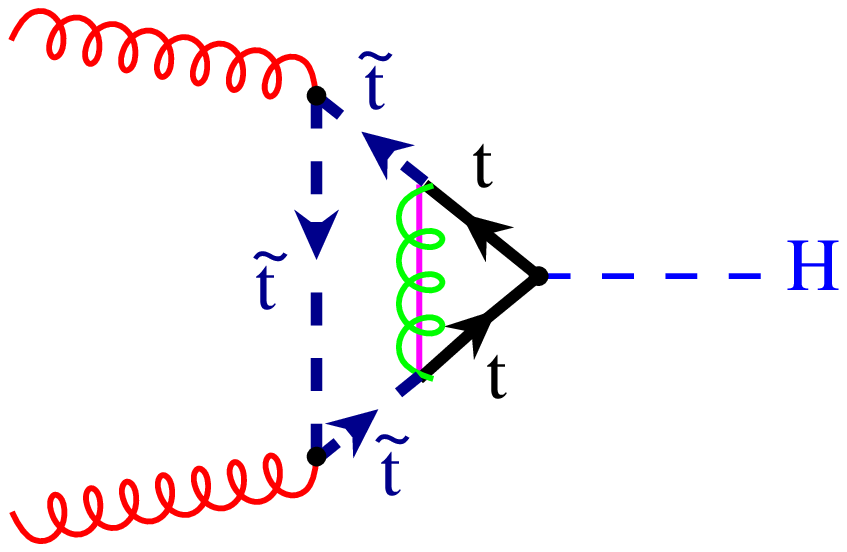}} &
      \epsfxsize=9em
      \raisebox{0em}{\epsffile[120 500 380 680]{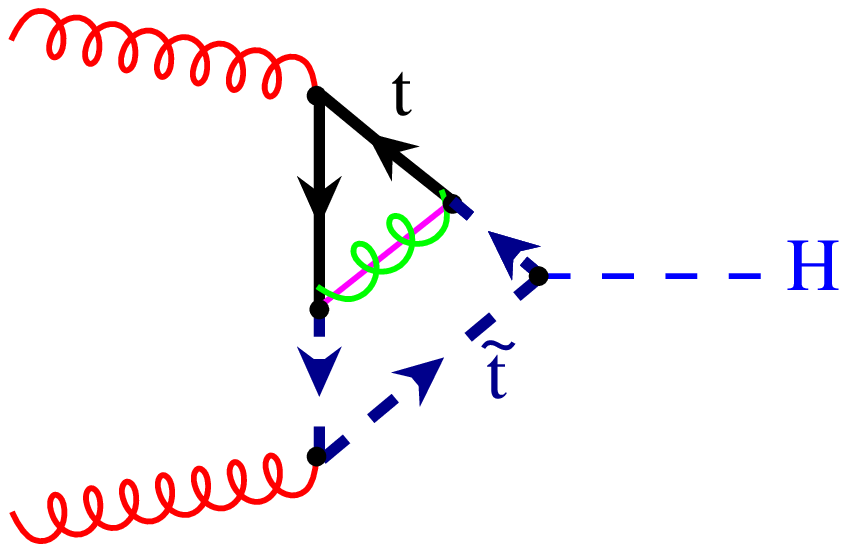}} \\
      $(a)$ & $(b)$ & $(c)$ & $(d)$
    \end{tabular}
    \parbox{14.cm}{
      \caption[]{\label{fig::vertnlo}\sloppy
        Sample two-loop Feynman diagrams contributing to gluon
        fusion in the \mssm{}.  
        Apart from the pure {\abbrev QCD} corrections to the
        top quark $(a)$ and top squark $(b)$ triangle, there are also
        corrections involving the gluino $(c)$, $(d)$.  }}
  \end{center}
\end{figure}

In the \sm{}, the coefficient function $C_1$ is known to order
$\alpha_s^4$.  The $\alpha_s^2$-terms have been computed in
Refs.~\cite{Inami:1982xt,Djouadi:1991tk}, while the contributions of
order $\alpha_s^3$ were obtained in
Refs.~\cite{Chetyrkin:1997iv,Chetyrkin:1997un} (see also
\reference{Kramer:1996iq}), and the terms of order $\alpha_s^4$ in
\reference{Chetyrkin:1997un}. For later reference, let us quote the
\nlo{} result at this point:
\begin{equation}
\begin{split}
  C_1^{\rm SM}(\alpha_s) &= -\frac{\alpha_s}{3\pi}\left( 1 +
    \frac{11}{4} \api \right) + \order{\alpha_s^3} \,.
\end{split}
\end{equation}
Within the \sm{} framework, there exists an all-order low-energy
theorem~\cite{Chetyrkin:1997un} which expresses $C_1^{\rm SM}$ in terms
of $\beta(\alpha_s)$ and $\gamma_m(\alpha_s)$, the anomalous dimensions
of $\alpha_s$ and $\Mtop{}$, respectively.  The result of
\reference{Dawson:1996xz} for the squark-contribution to the effective
theory at order $\alpha_s^2$ was derived by replacing the
$\beta$-function with its pure squark contribution, and $\gamma_m$ with
the anomalous dimension of the squark mass. The obtained expression was
multiplied by a factor $\propto (M_t/\Msusy{t})^2$, unaffected by
renormalization.\footnote{Assuming this (non-supersymmetric) theory of
  scalar quarks, we reproduced the result of \reference{Dawson:1996xz}
  in our diagrammatic approach. Only gluonic corrections to
  \fig{fig::vertlo}\,$(b)$ contribute in this case, {\it e.g.}\ 
  \fig{fig::vertnlo}\,$(b)$.}

Our explicit calculation shows that for the correct effective Higgs-gluon
coupling, the top squark contribution can not be considered on its own.
In fact, as soon as the proper renormalization of the Higgs-squark
coupling is taken into account, the pure squark contribution receives
ultra-violet divergences which are canceled by a contribution
coming from top quarks and gluinos, like the ones in
\fig{fig::vertnlo}\,$(c)$,\,$(d)$.
As a consequence of this important conceptual difference, also the
numerical results for hadronic Higgs production have to be
re-analyzed.

In this paper, we follow the diagrammatic approach of
Refs.~\cite{Chetyrkin:1997iv,Chetyrkin:1997un} in order to evaluate the
effective Higgs-gluon interaction in the \mssm{}, including terms of
order $\alpha_s^2$.  As explained above, this involves three types of
two-loop Feynman diagrams:
Gluonic corrections to top quark loops, {\it e.g.}\ \fig{fig::vertnlo}\,$(a)$;
gluonic corrections to top squark loops, {\it e.g.}\
  \fig{fig::vertnlo}\,$(b)$; gluino-contributions, {\it e.g.}\
  \fig{fig::vertnlo}\,$(c)$,\,$(d)$.
The resulting effective Lagrangian can then be used to evaluate the
Higgs production rate at a hadron collider to \nlo{}.

\subsection{Calculational details}

It is well known that Dimensional Regularization
(\dreg{})~\cite{'tHooft:fi} breaks supersymmetry.  This can be cured by
adding finite counterterms that restore the supersymmetric Ward
identities.  Alternatively, one can choose a regularization procedure
which preserves supersymmetry. A viable method which is very similar to
\dreg{} is Dimensional Reduction (\dred{})~\cite{Siegel:1979wq} (see
also~\cite{Jack:1997sr}).  The essential difference to \dreg{} is that
the tensor algebra is performed in four instead of $D=4-2\epsilon$
dimensions, whereas the loop integrals are still $D$ dimensional.  We
checked that the final result is the same, both in \dred{} and in
\dreg{}, provided the aforementioned finite counterterms are taken into
account.

The generation and evaluation of the Feynman diagrams proceeds fully
automatically through the chain
\begin{center}
\begin{tabular}{ccccccc}
{\tt QGRAF} &$\to$& 
{\tt Q2E} &$\to$&
{\tt EXP} &$\to$&
{\tt MATAD}/{\tt MINCER} \\
\cite{Nogueira:1993ex} &&
\cite{q2e} &&
\cite{exp} &&
\cite{Steinhauser:2000ry}/\cite{Larin:1991fz} \\
generation &&
analyzation &&
\begin{tabular}{c}
asymptotic \\ expansion
\end{tabular}
&&
calculation
\end{tabular}
\end{center}
{\tt Q2E} translates the output of {\tt QGRAF} into a format that is
suitable for further manipulation. {\tt EXP} applies asymptotic
expansions on a diagrammatical level, according to a user-defined
hierarchy of mass (possibly also momentum) scales, and produces output
that can be given to the integration packages {\tt MINCER} and {\tt
  MATAD} immediately. The latter use the {\it
  Integration-by-Parts}
algorithm~\cite{Chetyrkin:1981qh} in order to evaluate
the single-scale integrals analytically.

\subsection{The effective Lagrangian at next-to-leading order}

Various methods for the evaluation of $C_1$ are described in
Refs.~\cite{Chetyrkin:1997un,Steinhauser:2002rq}. We follow the most
direct one, which means to apply the projector
\begin{equation}
\begin{split}
P_{\mu\nu}^{ab}(p_1,p_2) = \delta^{ab}\,
\frac{p_1\cdot p_2\,g_{\mu\nu} - p_{1\nu} p_{2\mu} - p_{1\mu} p_{2\nu}}
{8\,(D-2)(p_1\cdot p_2)^2}
\end{split}
\end{equation}
to the $gg\phi$ vertex diagrams (sample diagrams are shown in
Figs.~\ref{fig::vertlo} and~\ref{fig::vertnlo}), and evaluating them at $p_1 =
p_2=0$.  Here, $\mu$, $\nu$ and $a$, $b$ are the Lorentz and color
indices of the external gluons, and $p_1$, $p_2$ are their momenta. This
will result in the product $\zeta_3^0\,C_1^0$, where $\zeta_3^0$ is the
bare decoupling constant of the gluon field obtained from the one-loop
gluon propagator involving top quarks, top squarks, and
gluinos:\footnote{We refrain from quoting terms proportional to
  $\gamma_{\rm E}$ and $\ln 4\pi$ that drop out of
  $\overline{\mbox{\scriptsize DR}}$-renormalized quantities.}
\begin{equation}
\begin{split}
\zeta_3^0 &= 1 + \api \bigg(
\frac{3}{4\ep} + L(\ep) \bigg) + \order{\alpha_s^2}\,,
\qquad \mbox{with}\\
L(\ep) &= \frac{1}{24}\left( 12\,\Lgluino +  \LstopL + \LstopR
   + 4\,\Ltop \right)
+
\frac{\ep}{12}\left(
 3\,\Lgluino^2 + \frac{1}{4}\left(\LstopL^2 + \LstopR^2\right) 
 + \Ltop^2 + \frac{9}{2}\,\zeta_2 \right)\,,
\label{eq::zeta30}
\end{split}
\end{equation}
\begin{equation}
\begin{split}
\Ltop &= \ln\frac{\mu^2}{\Mtop^2}\,,\qquad
\LstopL = \ln\frac{\mu^2}{\Msusy{L}^2}\,,\qquad
\LstopR = \ln\frac{\mu^2}{\Msusy{R}^2}\,,\qquad
\Lgluino = \ln\frac{\mu^2}{\Mgluino^2}\,,
\end{split}
\end{equation}
and $\zeta_2 \equiv \pi^2/6$. $\Msusy{L/R}$ is the mass for the
supersymmetric partner of the left/right-handed top quark, $\Mgluino$ is
the gluino mass, and $\Mtop$ the top quark mass.  One-loop
renormalization of the strong coupling constant and the masses (see,
{\it e.g.}, \reference{Kraml:1999qd}) leads to $C_1(\tilde \alpha_s)$,
where $\tilde\alpha_s$ is the renormalized coupling constant in the full
theory. It is related to $\alpha_s$, the coupling in standard five-flavor
{\abbrev QCD} (see \eqn{eq::defalphas}), by
\begin{equation}
\begin{split}
\alpha_s = (\zeta_g)^2\,\tilde \alpha_s\,,\qquad
\mbox{where}\qquad
\zeta_g^2 = 1 - \api L(\ep) + \order{\alpha_s^2}\,.
\label{eq::zetag}
\end{split}
\end{equation}
$L(\ep)$ is defined in \eqn{eq::zeta30}.

For brevity, we will focus on the following two cases in this letter:
\begin{equation}
\begin{split}
(A)\qquad & 
\Mgluino{}\gg \Msusy{L}=\Msusy{R}\equiv \Mstop{} \sim \Mtop{}\,,\\
(B)\qquad & 
\Msusy{R} \gg \Mgluino{}\gg \Msusy{L} \sim \Mtop{}\,,
\label{eq::cases}
\end{split}
\end{equation}
These cases are motivated by the fact that the supersymmetric (\susy{})
contributions to the gluon-Higgs coupling are proportional to
$\Mtop{}^2/\Mstop{}^2$.  Sizable effects are thus only expected if at
least one of the top squarks has a mass of the order of the top quark
mass.  Bottom squark contributions are suppressed by
$(M_b/\Msusy{b})^2\cdot\tan\beta$ and can savely be neglected for
not-too-large values of $\tan\beta$, as we assume them in this paper.
Furthermore, as already mentioned in the Introduction, we do not
consider the mixing of the top squarks.

The one-loop calculation of $C_1$ involves only diagrams with a single
mass scale: $\Mstop{}$ or $\Mtop{}$. At two-loop order, up to three
different masses can appear in the loop diagrams (see, e.g.,
\fig{fig::vertnlo}\,$(c)$,\,$(d)$), which makes the calculation quite
tedious.  However, adopting the hierarchies of the cases $(A)$ and $(B)$
introduced above, one can apply asymptotic expansions~\cite{Smi02} (see
also Ref.~\cite{Harlander:1998dq} for pedagogic examples) in order to reduce
the multi-scale to single-scale Feynman diagrams. This leads to
significantly simpler integrals and final results of a handy structure
(powers and logarithms of the masses).

For later convenience, we parametrize the \susy{} contributions to the
coefficient function in the following way:
\begin{equation}
\begin{split}
  C_1 &= g_t^\phi\,C_1^{\rm SM}\,c_0(\MtL,\MtR)\left[ 1 + \api\,c^{\rm
      SUSY} \right]\,,\qquad \mbox{where}
\end{split}
\end{equation}
\begin{equation}
\begin{split}
c_0(\MtL,\MtR) &= 1 + \frac{1}{4}\left( \MtL + \MtR \right)\,,\qquad
\MtL = \frac{\Mtop^2}{\Msusy{L}^2}\,,\qquad
\MtR = \frac{\Mtop^2}{\Msusy{R}^2}\,.
\label{eq::c0}
\end{split}
\end{equation}
$g_t^\phi$ represents the coupling of the Higgs boson to top quarks
which is specified below.
We evaluate case $(A)$ of \eqn{eq::cases} by first considering the
following subcase:
\begin{equation}
\begin{split}
(A^\prime)\qquad & 
\Mgluino{}\gg \Msusy{L}=\Msusy{R}\equiv \Mstop{} \gg \Mtop{}\,.
\end{split}
\end{equation}

Expressing the top quark and squark mass in the
on-shell scheme, one gets:
\begin{equation}
\begin{split} 
c^{\rm SUSY}[A'] &=
\frac{\Mts}{c_0(\Mts,\Mts)}\bigg[
a^{(0)} + \Mtg\,a^{(1)} + \Mtg^2\,a^{(2)} + \order{\Mtg^3} \bigg] +
\order{\alpha_s}\,,
\end{split}
\end{equation}
\begin{equation}
\begin{split}
a^{(0)} &= \frac{11}{8} 
  + \frac{1}{3}\Lgs + \Ltg\,,\qquad
a^{(1)} = \frac{1}{18} 
  - \frac{3}{4}\Lgs 
  + \frac{11}{12}\Ltg
+ \Mts\,\left(\frac{14}{9} + \frac{4}{3}\Ltg \right)\,,\\
a^{(2)} &=
 -\frac{157}{36} + \zeta_2 - \frac{3}{4}\Lgs - \frac{7}{12}\Ltg
- \frac{1}{2}\Lgs\Ltg  \\&\quad
+ \Mts \left(-\frac{37}{9} + \zeta_2- \frac{13}{12}\Lgs - \frac{7}{12}\Ltg
- \frac{1}{2}\Lgs\Ltg \right)
+ \Mts^2\,\left(\frac{25}{12} + 3\,\Ltg\right)\,,
\label{eq::(A)}
\end{split}
\end{equation}
with 
\begin{equation}
\begin{split}
\Mts & =\frac{\Mtop^2}{\Mstop^2}\,, \qquad
\Mtg=\frac{\Mstop^2}{\Mgluino^2}\,,
\qquad
\Ltg=\ln\frac{\Mtop^2}{\Mgluino^2}\,,\qquad 
\Lgs=\ln\frac{\Mgluino^2}{\Mstop^2}\,.
\end{split}
\end{equation}
In the numerical analysis of Section~\ref{sec::num},
the coefficients $a^{(i)}$ are included up to $i=6$.
Note that the explicit $\mu$ dependence drops out 
which has to be the case as the anomalous dimension of $C_1$ has to cancel the
$\mu$ dependence of $\alpha_s$ in the one-loop term
(cf. Eq.~(\ref{eq::c1ren})).
This is a strong check for our calculation.

An observation concerning \eqn{eq::(A)} is that the two-loop
contribution of order $\Mtop{}^2/M_{\tilde{t}}^2$ depends
logarithmically on $\Mgluino{}$
which is a consequence of the simultaneous decoupling of all heavy 
particles.\footnote{Note that the 
  Appelquist-Carazzone decoupling theorem~\cite{Appelquist:tg} 
  is not applicable here
  since removing the gluino from the \mssm{} leads to a non-renormalizable
  theory, if the supersymmetric relation between the top and the stop Yukawa
  coupling should be preserved.}
However, in the limit that {\it all} \susy{} particles are heavy, as expected,
the \sm{} result is recovered.  Let us add that the 
bare two-loop result even has contributions $\propto \Mgluino{}^2\cdot
\Mtop{}^2/\Mstop{}^2$, but those are canceled by the on-shell counter
term of the squark mass.

Another remarkable observation concerning \eqn{eq::(A)} is that higher
order terms in $\Mtop{}^2/\Mstop^2$ identically vanish, {\it i.e.}, the
coefficients $a^{(i)}$, $i=0,1,2,\ldots$ are exact.  This
means that \eqn{eq::(A)} is valid for arbitrary values of $\Mtop{}$ and
$\Mstop{}$, provided that $\Mgluino{}$ is sufficiently large. In
particular, it covers case $(A)$ of \eqn{eq::cases}:
\begin{equation}
\begin{split}
c^{\rm SUSY}[A] \equiv c^{\rm SUSY}[A']\,.
\end{split}
\end{equation}
This equality has been checked up to order $1/\Mgluino^{16}$, and it is
suggestive that it holds to all orders in the inverse gluino mass.

\begin{figure}
  \begin{center}
    \leavevmode
    \begin{tabular}{cc}
      \epsfxsize=17em
      \epsffile[70 250 540 550]{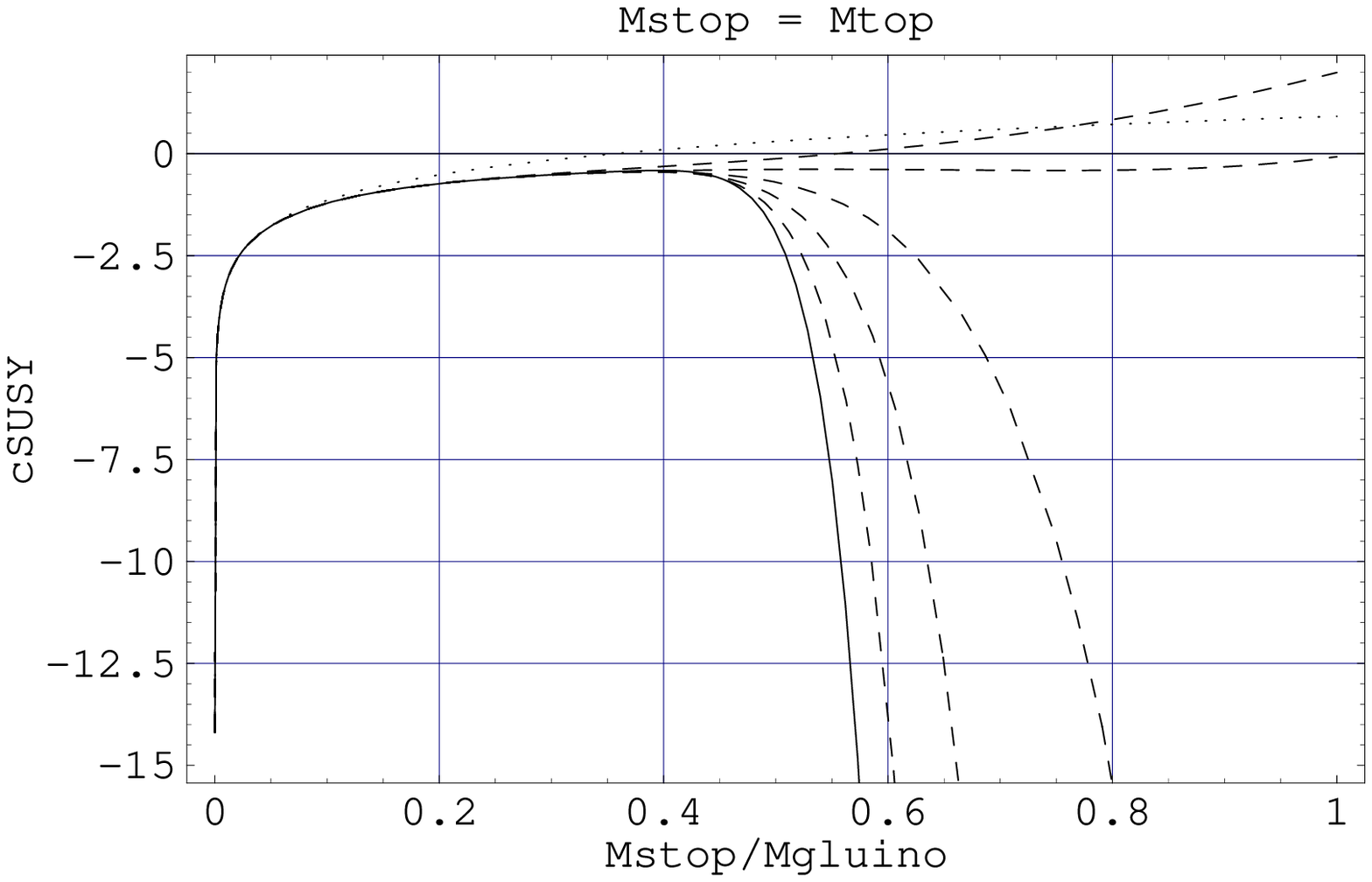} &
      \epsfxsize=17em
      \epsffile[70 250 540 550]{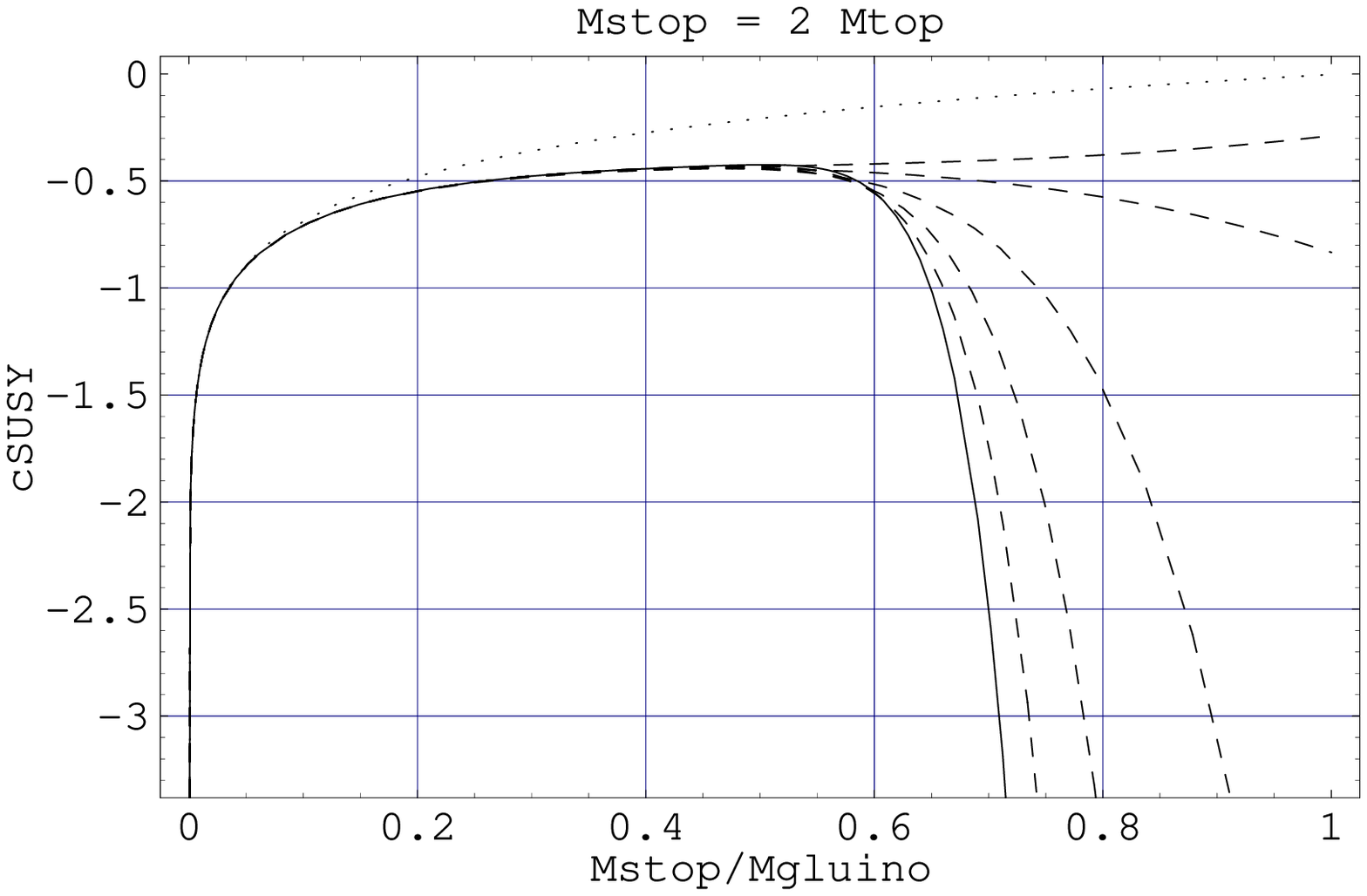}\\
      $(a)$ & $(b)$
    \end{tabular}
    \parbox{14.cm}{
      \caption[]{\label{fig::(A)}\sloppy
        $c^{\rm SUSY}$ for
        Case $(A)$ of \eqn{eq::cases} with $(a)$ $\Mstop{} = \Mtop{}$
        and $(b)$ $\Mstop{} = 2\Mtop{}$, as a function of the ratio
        $\Mstop{}/\Mgluino{}$.  The individual lines correspond to
        different orders of the expansion in the square of this ratio.
        The dotted line includes only the leading term, whereas the solid
        line incorporates corrections up to order $(\Mstop/\Mgluino)^{12}$.}}
  \end{center}
\end{figure}

In order to quantify the condition $\Mgluino{}\gg (\Mtop{},\Mstop{})$,
we show in \fig{fig::(A)} the expansion of \eqn{eq::(A)}, including
successively higher orders in 
$x_{sg}=\Mstop^2/\Mgluino^2$, for fixed values of $x_{ts}$.
One clearly observes that the expansion breaks down dramatically beyond
a certain value of $x_{sg}$.  However, we conclude that for $\Mgluino
\gtrsim 2.5\,\Mstop$ and $\Mstop\geq \Mtop$, the expansion should be
accurate to better than $1\%$.  Note also that, as mentined above, the
expression diverges logarithmically like $\ln(x_{sg})$ as $x_{sg}\to 0$.

As an interesting special case, let us consider the limit
$\Msusy{L}=\Msusy{R}=\Mtop{}$ which leads to the compact formula
\begin{equation}
\begin{split}
C_1[ \Msusy{L} = \Msusy{R} = \Mtop ] &= -g_t^\phi\frac{\alpha_s}{2\pi}\left[ 
  1 + \api\left( \frac{11}{3} + \frac{4}{9}\Ltg \right) \right]
+\order{\frac{\Mtop^2}{\Mgluino^2},\alpha_s^3}
\end{split}
\end{equation}
It is worth noting that the coefficient $a^{(0)}$ in Eq.~(\ref{eq::(A)})
is positive only for $\Mgluino<e^{33/32}\Mtop\approx 2.8\Mtop$.

Considering case $(B)$ of \eqn{eq::cases}, we proceed in complete
analogy to case $(A)$, {\it i.e.}, we first evaluate the subcase
\begin{equation}
\begin{split}
(B^\prime)\qquad 
\Msusy{R} \gg \Mgluino{}\gg \Msusy{L} \gg \Mtop{}\,.
\end{split}
\end{equation}

Keeping only the leading term in $1/\Msusy{R}$ at \nlo{}, we find
\begin{equation}
\begin{split}
c^{\rm SUSY}[B^\prime] &= \frac{\MtL}{c_0(\MtL,\MtR)}\,\bigg[
 b^{(0)} + \MLg\,b^{(1)} + \MLg^2\,b^{(2)}
+ \order{\MtR,\MLg^3}\bigg] + \order{\alpha_s}\,,
\end{split}
\end{equation}
\begin{equation}
\begin{split}
b^{(0)} &= \frac{29}{48} + \frac{\LgR}{12} + \frac{\LgL}{6} +
 \frac{\Ltg}{2}\,,\\
b^{(1)} &= -\frac{1}{18} - \frac{5}{24}\LgL
+ \frac{11}{24}\Ltg + \MtL ( \frac{13}{18} + \frac{2}{3}\Ltg )\,,\\
b^{(2)} &= 
- \frac{163}{72} + \frac{1}{2}\zeta_2 - \frac{1}{8}\LgL - \frac{7}{24}\Ltg -
\frac{1}{4}\LgL\Ltg 
\\&
\quad+ \MtL \left(-\frac{85}{36} + \frac{1}{2}\zeta_2
  - \frac{3}{8}\LgL - \frac{7}{24}\Ltg 
  - \frac{1}{4}\LgL\Ltg \right)
+ \MtL^2  \left(\frac{49}{48} + \frac{3}{2}\Ltg\right)\,,
\label{eq::(B)}
\end{split}
\end{equation}
where
\begin{equation}
\begin{split}
\MtL&=\frac{\Mtop^2}{\Msusy{L}^2}\,,\qquad 
\MtR=\frac{\Mtop^2}{\Msusy{R}^2}\,,\qquad 
\MLg=\frac{\Msusy{L}^2}{\Mgluino^2}\,,\\
\Ltg&=\ln\frac{\Mtop^2}{\Mgluino^2}\,,\qquad
\LgL=\ln\frac{\Mgluino^2}{\Msusy{L}^2}\,,\qquad
\LgR=\ln\frac{\Mgluino^2}{\Msusy{R}^2}\,.
\end{split}
\end{equation}
We again evaluated the coefficients up to order $\MtR^6$ and $\MLg^6$
and use them in the numerical analyses below.
Note that this result still has a logarithmic dependence on $\Msusy{R}$;
this comes in addition to the logarithmic dependence on
$\Mgluino$ as $\Mgluino\to \infty$, as it was already observed in case~$(A)$.

Also in analogy to case~$(A)$, the coefficients $b^{(i)}$,
$i=0,1,2,\ldots$ are exact, meaning that \eqn{eq::(B)} also covers case~$(B)$:
\begin{equation}
\begin{split}
c^{\rm SUSY}[B] = c^{\rm SUSY}[B']\,.
\end{split}
\end{equation}
For sufficiently large values of $\Msusy{R}$,
an analogous study as in case $(A)$ to quantify the condition
$\Mgluino\gg (\Msusy{L},\Mtop)$ leads to very similar conclusions, so
that we refrain from presenting it here.

\section{\label{sec::num}Higgs Decay Rate and Production Cross Section}

\subsection{Decay rate}

From Eq.~(\ref{eq::leff}) one can derive a general expression for the
inclusive $\phi\to gg$ decay width,\footnote{Note that at {\scriptsize
    NLO} also the final states $ggg$ and $q\bar qg$ contribute to this
  quantity.}
\begin{equation}
\Gamma(\phi\to gg)
=\Gamma_0\left(1+\delta\right)
\,,
\end{equation}
$\Gamma_0$ represents the complete
\lo{} result which is given by (see, {\it e.g.}, Ref.~\cite{Spira:1997dg})
\begin{equation}
\begin{split}
  \Gamma_0 &=
  \frac{G_F\alpha_s^2 M_\phi^3}{16\sqrt{2}\pi^3}
  \left| \sum_{q\in\{b,t\}} g_q^\higgs\,\left(A(\tau_q) +
      \frac{M_q^2}{\Msusy{qL}^2}\,\tilde A(\tilde
      \tau_{qL})
      + \frac{M_q^2}{\Msusy{qR}^2}\,\tilde A(\tilde
      \tau_{qR})\right)
  \right|^2\,,\\
  \tau_q &=
  \frac{4M_q^2}{M_\higgs^2}\,,\quad
  \tilde\tau_{qL} = \frac{4\Msusy{qL}^2}{M_\higgs^2}\,,\quad
  \tilde\tau_{qR} = \frac{4\Msusy{qR}^2}{M_\higgs^2}\,,\quad
  \\
  A(\tau) &= \tau\left[1 + (1-\tau)\,f(\tau)\right]\,,\qquad 
  \tilde A(\tilde\tau) = -\frac{1}{2}\,\tilde\tau\left[1 -
    \tilde\tau\,f(\tilde\tau)\right]\,,
  \label{eq::gamma0}
\end{split}
\end{equation}
with
\begin{equation}
\begin{split}
f(\tau) &= \left\{
\begin{array}{ll}
\arcsin^2\left(\frac{1}{\sqrt{\tau}}\right)\,, &
\tau\geq 1\,,\\
-\frac{1}{4}\left[
  \ln\frac{1 + \sqrt{1-\tau}}{1-\sqrt{1-\tau}} - i\pi\right]^2\,,&
\tau<1\,.
\end{array}
\right.
\label{eq::tildeA}
\end{split}
\end{equation}
Relative to the \sm{} case, the top quark coupling reads
\begin{equation}
\begin{split}
g_t^h = \frac{\cos\alpha}{\sin\beta}\,,\qquad
g_t^H = \frac{\sin\alpha}{\sin\beta}\,,
\end{split}
\end{equation}
where $h,H$ denote the light and heavy neutral scalar Higgs boson,
respectively, and $\alpha$ is the mixing angle between weak and mass
eigenstates in the Higgs sector. Since our focus is on not-too large
$\tan\beta$, we will neglect the effect of bottom quarks and bottom
squarks throughout the paper. In this limit, $g_t^\phi$, and therefore
also $\tan\beta$, enters our expression as an overall factor.

At \nlo{}, the correction term can be written as
\begin{equation}
\begin{split}
  \delta &= \api\left( \delta^{\rm SM} + 2\,c^{\rm SUSY} \right)\,,
\end{split}
\end{equation}
\begin{equation}
\begin{split}
\mbox{with}\qquad  \delta^{\rm SM} \,\,=\,\, 
    \frac{95}{4} - \frac{7}{6}n_l
    + \left(\frac{11}{2} - \frac{1}{3}
  n_l\right)\ln\frac{\mu^2}{M_\phi^2}\,,
\label{eq::decay}
\end{split}
\end{equation}
where $n_l=5$ is the number of light quark flavors and $c^{\rm
  SUSY}$ is given in Eqs.~(\ref{eq::(A)}) and~(\ref{eq::(B)}).

\begin{figure}
  \begin{center}
    \leavevmode
    \begin{tabular}{c}
      \epsfxsize=26em
      \epsffile[80 250 540 550]{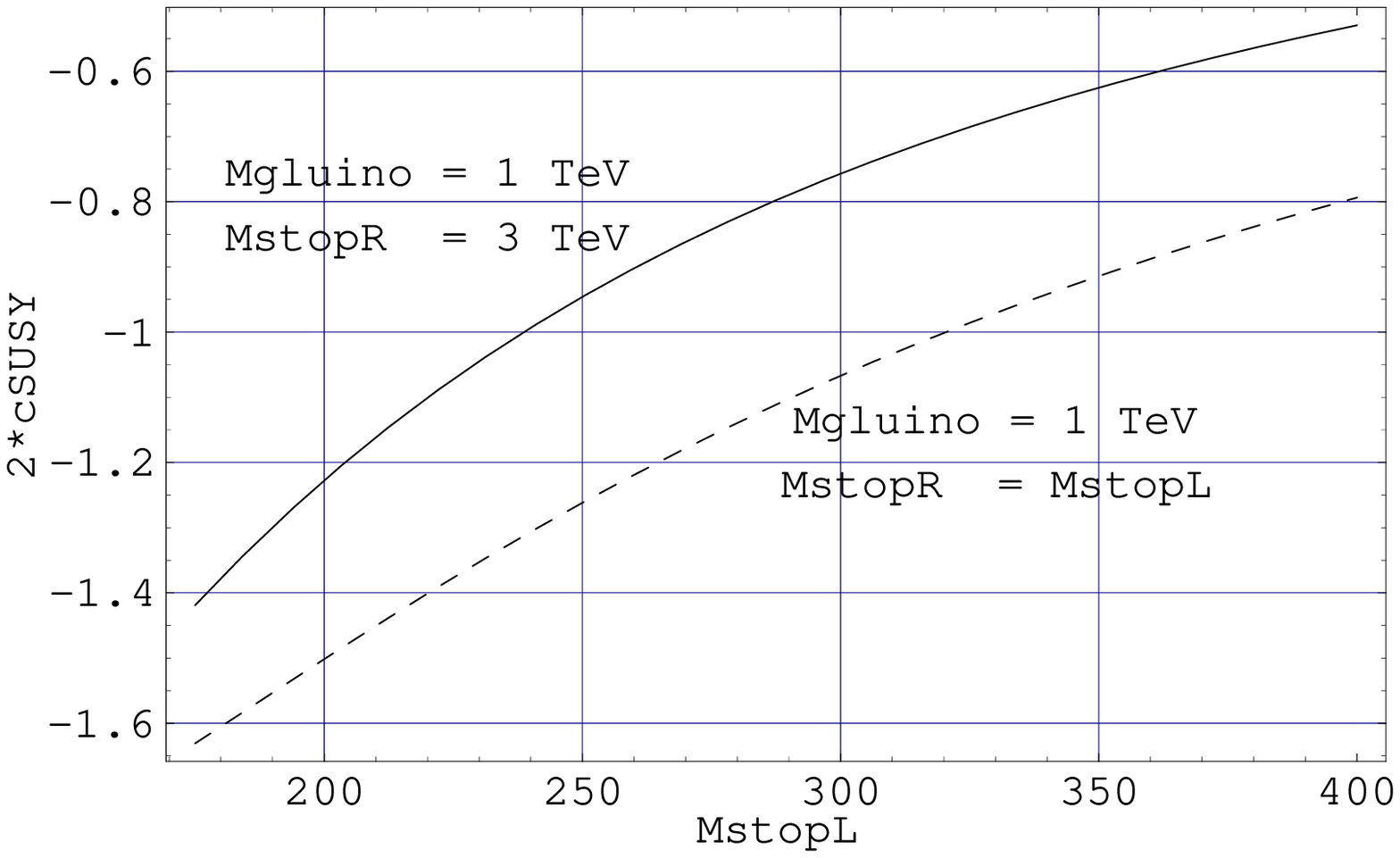}
    \end{tabular}
    \parbox{14.cm}{
      \caption[]{\label{fig::csusy}\sloppy
        The quantity $2c^{\rm SUSY}$ as a function of the squark mass
        for $\Mgluino=1$\,TeV and $\Mtop=175$\,GeV. Notice that the
        \susy{} effects are {\it negative}.
        }}
  \end{center}
\end{figure}

The effect of the \susy{} corrections is shown in Fig.~\ref{fig::csusy}
where the quantity $2 c^{\rm SUSY}$ is plotted as a function of the
squark mass for $\Mgluino=1$\,TeV.
For $\alpha_s=0.1$, we get corrections of the order of $-5\%$.

\subsection{Production cross section}

\subsubsection{Partonic Results}
The hadronic cross section $\sigma_{hk} \equiv \sigma(hk\to \higgs+X)$ for
Higgs production can be written as
\begin{equation}
\begin{split}
  \sigma_{hk}(z) = \sum_{i,j}
  \int_z^1\dd x_1 \int_{z/x_1}^1
  \dd x_2\,\,
  \varphi_{i/h}(x_1)\,
  \varphi_{j/k}(x_2)\,
  \hat\sigma_{ij}\left(\frac{z}{x_1x_2}\right)\,,\qquad
  z\equiv \frac{M_\higgs^2}{s}\,,
\end{split}
\end{equation}
where $i$, $j$ denote any partons inside the hadrons $h$, $k$, and
$\varphi(x)$ is a parton density. $\hat\sigma_{ij}$ is the partonic
cross section, and $s$ is the hadronic c.m.\,energy

It is convenient to write this partonic cross section as
\begin{equation}
\begin{split}
\hat \sigma_{ij}(x) = \sigma_0\,\Delta_{ij}(x)\,,\qquad
x = \frac{M_\higgs^2}{\hat s}\,,\qquad
\sigma_0 &= \frac{\pi^2}{8M_\higgs^3}\Gamma_0\,,
\end{split}
\end{equation}
where $\Gamma_0$ is defined in \eqn{eq::gamma0}, $\hat s$ is the
partonic center-of-mass energy, and
\begin{equation}
\begin{split}
  \Delta_{ij}(x) = \delta_{ig}\delta_{jg}\,\delta(1-x) +
  \api\,\Delta_{ij}^{(1)}(x) + \order{\alpha_s^2}\,.
\end{split}
\end{equation}

We find for the \nlo{} terms ($q\in \{u,d,s,c\}$)\footnote{
  The renormalization and factorization scales
  have been identified with $M_\phi$ and
  can easily be reconstructed from lower order results.}: 
\begin{equation}
\begin{split}
  \Delta^{(1)}_{gg} &=  \Delta^{(1),\rm SM}_{gg} + 2\,c^{\rm
    SUSY}\,\delta(1-x)\,,\\[1em]
  \Delta^{(1),\rm SM}_{gg} &= \left ( \frac{11}{2} + 6 \zeta_2 \right )
  \delta(1-x) + 12 \left [ \frac{\ln(1-x)}{1-x} \right ]_+ -
  12x(-x+x^2+2)\ln(1-x)
  \\
  & \quad -\frac{6 (x^2+1-x)^2}{1-x}\ln(x) -\frac{11}{2} (1-x)^3
  \,, \\
  \Delta^{(1)}_{qg} &= \Delta^{(1),\rm SM}_{qg} = -\frac{2}{3} \left (
    1+(1-x)^2 \right )\ln \frac{x}{(1-x)^2}-1+2x
  -\frac{1}{3}x^2  \,, \\
  \Delta^{(1)}_{q\bar q} &= \Delta^{(1),\rm SM}_{q\bar q} =
  \frac{32}{27}(1-x)^3 \,,
\end{split}
\end{equation}
where $c^{\rm SUSY}$ is defined in \eqn{eq::decay} and the subscript ``+''
  denotes the standard plus distribution.  $\Delta^{\rm
  SM}_{ij}(x)$ is the \nlo{} Standard Model
expression~\cite{Dawson:1990zj,Djouadi:1991tk}. All other partonic
subprocesses vanish at \nlo{}.  Note that the \susy{} corrections only
modify the coefficient of the $\delta(1-x)$ contribution in the gluonic
channel.

\subsubsection{Hadronic Results}

\begin{figure}
  \begin{center}
    \leavevmode
    \begin{tabular}{c}
      \epsfxsize=20em
      \epsffile[110 265 465 560]{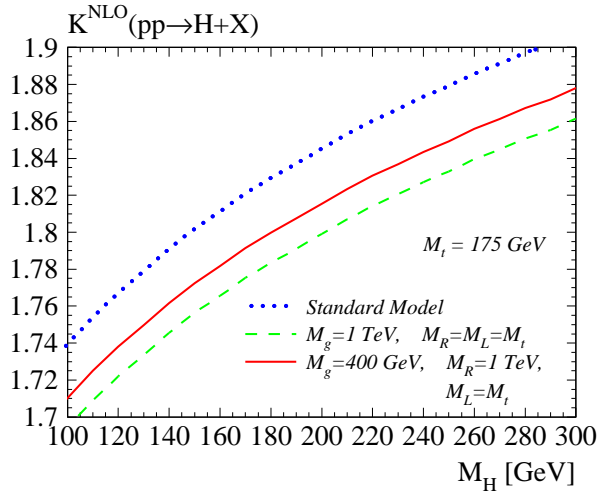}\\
    \end{tabular}
    \parbox{14.cm}{
      \caption[]{\label{fig::ksusy1}\sloppy
        K-factors at \nlo{} for Higgs production in gluon fusion in the
        Standard Model (dotted) and in Supersymmetry. Dashed line:
        $\Msusy{R} = \Msusy{L} = \Mtop = 175$\,GeV, $\Mgluino = 1$\,TeV
        --- Solid line: $\Msusy{R} = 1$\,TeV, $\Msusy{L} = \Mtop =
        175$\,GeV, $\Mgluino = 400$\,GeV. The bottom Yukawa coupling has
        been neglected. We use the \lo{} and \nlo{} parton densities of
        {\tt MRST2001}~\cite{Martin:2001es}. }}
  \end{center}
\end{figure}

\fig{fig::ksusy1} shows the \nlo{} K-factor for hadronic Higgs production
in the Standard Model and in Supersymmetry, for two different choices of
squark and gluino masses, corresponding to the cases $(A)$ and $(B)$.
We notice that, even though the overall normalization of the cross
section in \susy{} can be a multiple of the \sm{} value, the {\abbrev QCD}
corrections are very similar in both cases, differing by less than 10\%.
However, in contrast to \reference{Dawson:1996xz}, we find that the
\susy{} effects are {\it negative} for large values of the gluino mass.

\section{Conclusions}

In this paper we considered the \nlo{} supersymmetric corrections to the
production and decay of a Higgs boson. We used the framework of an
effective Lagrangian where the heavy particles enter the coefficient
function, $C_1$, of the operator describing the gluon-Higgs coupling.
The practical calculation is based on the evaluation of the gluon-Higgs
vertex diagrams using asymptotic expansions. Our results are in
disagreement with \reference{Dawson:1996xz}, which is due to the reasons
that have been discussed above. The \susy{} corrections decrease the
effects from pure {\abbrev QCD} by less than 5\% in the considered
parameter space. A more detailed numerical analysis and the inclusion of
general squark mixing will be presented elsewhere.

\paragraph{Acknowledgments.}
We are grateful to K.~Chetyrkin, S.~Heinemeyer, G.~Kramer, A.~Penin, and
T.~Plehn for discussions, T.~Seidensticker for his help in the handling
of {\tt q2e/exp}, and G.~Weiglein for discussions and useful comments to
the manuscript.  We would like to thank the authors of
Ref.~\cite{Dawson:1996xz} for communications.


\begin{thebibliography}{99}



\def\app#1#2#3{{ Act.~Phys.~Pol.~}\jref{\bf B #1}{#2}{#3}}
\def\apa#1#2#3{{ Act.~Phys.~Austr.~}\jref{\bf#1}{#2}{#3}}
\def\annphys#1#2#3{{ Ann.~Phys.~}\jref{\bf #1}{#2}{#3}}
\def\cmp#1#2#3{{ Comm.~Math.~Phys.~}\jref{\bf #1}{#2}{#3}}
\def\cpc#1#2#3{{ Comp.~Phys.~Commun.~}\jref{\bf #1}{#2}{#3}}
\def\epjc#1#2#3{{ Eur.\ Phys.\ J.\ }\jref{\bf C #1}{#2}{#3}}
\def\fortp#1#2#3{{ Fortschr.~Phys.~}\jref{\bf#1}{#2}{#3}}
\def\ijmpc#1#2#3{{ Int.~J.~Mod.~Phys.~}\jref{\bf C #1}{#2}{#3}}
\def\ijmpa#1#2#3{{ Int.~J.~Mod.~Phys.~}\jref{\bf A #1}{#2}{#3}}
\def\jcp#1#2#3{{ J.~Comp.~Phys.~}\jref{\bf #1}{#2}{#3}}
\def\jetp#1#2#3{{ JETP~Lett.~}\jref{\bf #1}{#2}{#3}}
\def\jhep#1#2#3{{\small JHEP~}\jref{\bf #1}{#2}{#3}}
\def\mpl#1#2#3{{ Mod.~Phys.~Lett.~}\jref{\bf A #1}{#2}{#3}}
\def\nima#1#2#3{{ Nucl.~Inst.~Meth.~}\jref{\bf A #1}{#2}{#3}}
\def\npb#1#2#3{{ Nucl.~Phys.~}\jref{\bf B #1}{#2}{#3}}
\def\nca#1#2#3{{ Nuovo~Cim.~}\jref{\bf #1A}{#2}{#3}}
\def\plb#1#2#3{{ Phys.~Lett.~}\jref{\bf B #1}{#2}{#3}}
\def\prc#1#2#3{{ Phys.~Reports }\jref{\bf #1}{#2}{#3}}
\def\prd#1#2#3{{ Phys.~Rev.~}\jref{\bf D #1}{#2}{#3}}
\def\pR#1#2#3{{ Phys.~Rev.~}\jref{\bf #1}{#2}{#3}}
\def\prl#1#2#3{{ Phys.~Rev.~Lett.~}\jref{\bf #1}{#2}{#3}}
\def\pr#1#2#3{{ Phys.~Reports }\jref{\bf #1}{#2}{#3}}
\def\ptp#1#2#3{{ Prog.~Theor.~Phys.~}\jref{\bf #1}{#2}{#3}}
\def\ppnp#1#2#3{{ Prog.~Part.~Nucl.~Phys.~}\jref{\bf #1}{#2}{#3}}
\def\rmp#1#2#3{{ Rev.~Mod.~Phys.~}\jref{\bf #1}{#2}{#3}}
\def\sovnp#1#2#3{{ Sov.~J.~Nucl.~Phys.~}\jref{\bf #1}{#2}{#3}}
\def\sovus#1#2#3{{ Sov.~Phys.~Usp.~}\jref{\bf #1}{#2}{#3}}
\def\tmf#1#2#3{{ Teor.~Mat.~Fiz.~}\jref{\bf #1}{#2}{#3}}
\def\tmp#1#2#3{{ Theor.~Math.~Phys.~}\jref{\bf #1}{#2}{#3}}
\def\yadfiz#1#2#3{{ Yad.~Fiz.~}\jref{\bf #1}{#2}{#3}}
\def\zpc#1#2#3{{ Z.~Phys.~}\jref{\bf C #1}{#2}{#3}}
\def\ibid#1#2#3{{ibid.~}\jref{\bf #1}{#2}{#3}}

\newcommand{\jref}[3]{{\bf #1} (#2) #3}
\bibitem{Kniehl:1993ay}
B.~A.~Kniehl,
Phys.\ Rept.\  {\bf 240} (1994) 211.

\bibitem{Spira:1997dg}
M.~Spira,
\fortp{46}{1998}{203}.

\bibitem{Carena:2002es}
M.~Carena and H.E.~Haber,
Prog.\ Part.\ Nucl.\ Phys.\ {\bf 50} (2003) 63.

\bibitem{Graudenz:1992pv}
D.~Graudenz, M.~Spira, and P.M.~Zerwas,
\prl{70}{1993}{1372}.

\bibitem{Spira:1995rr}
M.~Spira, A.~Djouadi, D.~Graudenz, and P.M.~Zerwas,
\npb{453}{1995}{17}.

\bibitem{Dawson:1990zj}
S.~Dawson,
\npb{359}{1991}{283}.

\bibitem{Djouadi:1991tk}
A.~Djouadi, M.~Spira, and P.~M.~Zerwas,
Phys.\ Lett.\ B {\bf 264} (1991) 440.

\bibitem{Harlander:2002wh}
R.~V.~Harlander and W.~B.~Kilgore,
Phys.\ Rev.\ Lett.\  {\bf 88} (2002) 201801.

\bibitem{Anastasiou:2002yz}
C.~Anastasiou and K.~Melnikov,
Nucl.\ Phys.\ B {\bf 646} (2002) 220.

\bibitem{Ravindran:2003um}
V.~Ravindran, J.~Smith, and W.L.~van Neerven,
hep-ph/0302135.

\bibitem{Spira:1993bb}
M.~Spira, A.~Djouadi, D.~Graudenz, and P.M.~Zerwas,
Phys.\ Lett.\ B {\bf 318} (1993) 347.

\bibitem{Eichten:1984eu}
E.~Eichten, I.~Hinchliffe, K.~D.~Lane, and C.~Quigg,
Rev.\ Mod.\ Phys.\  {\bf 56} (1984) 579;\\
(E) {\bf 58} (1986) 1065.

\bibitem{Maltoni:2003pn}
F.~Maltoni, Z.~Sullivan, and S.~Willenbrock,
Phys.\ Rev.\ D {\bf 67} (2003) 093005

\bibitem{Harlander:2003ai} 
R.V.~Harlander and W.B.~Kilgore,
\prd{68}{2003}{013001}.


\bibitem{Dawson:1996xz}
S.~Dawson, A.~Djouadi, and M.~Spira,
Phys.\ Rev.\ Lett.\  {\bf 77} (1996) 16.

\bibitem{Spiridonov:br}
V.~P.~Spiridonov,
IYaI-P-0378.


\bibitem{Chetyrkin:1997un}
K.~G.~Chetyrkin, B.~A.~Kniehl, and M.~Steinhauser,
Nucl.\ Phys.\ B {\bf 510} (1998) 61.

\bibitem{Inami:1982xt}
T.~Inami, T.~Kubota, and Y.~Okada,
Z.\ Phys.\ C {\bf 18} (1983) 69.

\bibitem{Chetyrkin:1997iv}
K.~G.~Chetyrkin, B.~A.~Kniehl, and M.~Steinhauser,
Phys.\ Rev.\ Lett.\  {\bf 79} (1997) 353.

\bibitem{Kramer:1996iq}
M.~Kr\"amer, E.~Laenen, and M.~Spira,
Nucl.\ Phys.\ B {\bf 511} (1998) 523.

\bibitem{'tHooft:fi}
G.~'t Hooft and M.~J.~Veltman,
Nucl.\ Phys.\ B {\bf 44} (1972) 189.

\bibitem{Siegel:1979wq}
W.~Siegel,
Phys.\ Lett.\ B {\bf 84} (1979) 193.

\bibitem{Jack:1997sr}
I.~Jack and D.~R.~Jones,
hep-ph/9707278.


\bibitem{Nogueira:1993ex}
P.~Nogueira,
\jcp{105}{1993}{279}.

\bibitem{q2e}
T.~Seidensticker, unpublished.

\bibitem{exp}
T.~Seidensticker,
hep-ph/9905298;\\
R.~Harlander, T.~Seidensticker, and M.~Steinhauser,
Phys.\ Lett.\ B {\bf 426} (1998) 125.


\bibitem{Steinhauser:2000ry}
M.~Steinhauser,
\cpc{134}{2001}{335}.

\bibitem{Larin:1991fz}
S.A.~Larin, F.V.~Tkachov, and J.A.~Vermaseren,
NIKHEF-H-91-18.

\bibitem{Chetyrkin:1981qh}
K.G.~Chetyrkin and F.V.~Tkachov,
\npb{192}{1981}{159}.

\bibitem{Steinhauser:2002rq}
M.~Steinhauser,
\pr{364}{2002}{247}.


\bibitem{Kraml:1999qd}
S.~Kraml,
hep-ph/9903257;\\
H.~Eberl, Dissertation, TU Wien, 1998.



\bibitem{Smi02}
V.~A. Smirnov,
{\it Applied Asymptotic Expansions in Momenta and Masses},
(Springer-Verlag, Heidelberg, 2001).

\bibitem{Harlander:1998dq}
R.~Harlander and M.~Steinhauser,
Prog.\ Part.\ Nucl.\ Phys.\  {\bf 43} (1999) 167.

\bibitem{Appelquist:tg}
T.~Appelquist and J.~Carazzone,
Phys.\ Rev.\ D {\bf 11} (1975) 2856.


\bibitem{Martin:2001es}
A.~D.~Martin, R.~G.~Roberts, W.~J.~Stirling, and R.~S.~Thorne,
Eur.\ Phys.\ J.\ C {\bf 23} (2002) 73.


\end{thebibliography}
\end{document}